\documentclass[usegraphicx]{mn2e}
\usepackage{times}
\newlength{\colwidth}
\title[EROs in the UKIDSS UDS EDR]{Extremely red objects in the UKIDSS
Ultra Deep Survey Early Data Release}
\author[C.\ Simpson et al.]{Chris Simpson$^1$\thanks{E-mail:
cjs@astro.livjm.ac.uk}, 
Omar Almaini$^2$,
Michele Cirasuolo$^3$,
Jim Dunlop$^3$,
S\'{e}bastien Foucaud$^2$, \newauthor
Paul Hirst$^4$,
Rob Ivison$^{5,3}$,
Mat Page$^6$,
Steve Rawlings$^7$,
Kaz Sekiguchi$^8$,
Ian Smail$^9$, \newauthor
and Mike Watson$^{10}$\\
$^1$Astrophysics Research Institute, Liverpool John Moores University,
Twelve Quays House, Egerton Wharf, Birkenhead CH41 1LD\\
$^2$School of Physics and Astronomy, University of Nottingham,
University Park, Nottingham NG7 2RD\\
$^3$Institute for Astronomy, University of Edinburgh, Royal Observatory,
Blackford Hill, Edinburgh EH9~3HJ\\
$^4$Joint Astronomy Centre, 660 N.~A`oh\={o}k\={u} Place, Hilo, HI
96720, USA\\
$^5$UK Astronomy Technology Centre, Royal Observatory, Blackford Hill,
Edinburgh EH9~3HJ\\
$^6$Mullard Space Science Laboratory (MSSL), University College
London, Holmbury St. Mary, Dorking, Surrey RH5 6NT\\
$^7$Department of Physics, University of Oxford, Denys Wilkinson Building,
Keble Road, Oxford OX1 3RH\\
$^8$Subaru Telescope, National Astronomical Observatory of Japan, 650
N.~A`oh\={o}k\={u} Place, Hilo, HI 96720, USA\\
$^9$Institute for Computataional Cosmology, Department of Physics,
Durham University, South Road, Durham DH1 3LE\\
$^{10}$Department of Physics and Astronomy, University of Leicester,
Leicester LE1 7RH}
\begin{document}

\date{Version of \today}

\pagerange{\pageref{firstpage}--\pageref{lastpage}} \pubyear{2006}

\maketitle

\label{firstpage}

\begin{abstract}
We construct a sample of extremely red objects (EROs) within the
UKIDSS Ultra Deep Survey by combining the Early Data Release with
optical data from the Subaru/\textit{XMM-Newton\/} Deep Field. We find
a total of 3715 objects over 2013\,arcmin$^2$ with $R-K>5.3$ and
$K\leq20.3$, which is a higher surface density than found by previous
studies. This is partly due to our ability to use a small aperture in
which to measure colours, but is also the result of a genuine
overdensity of objects compared to other fields. We separate our
sample into passively-evolving and dusty star-forming galaxies using
their \textit{RJK\/} colours and investigate their radio properties
using a deep radio map. The dusty population has a higher fraction of
individually-detected radio sources and a higher mean radio flux
density among the undetected objects, but the passive population has a
higher fraction of bright radio sources, suggesting that AGNs are more
prevalent among the passive ERO population.
\end{abstract}


\section{Introduction}

Extremely Red Objects (EROs) are variously defined on the basis of an
optical--near-infrared colour, usually $R-K$ or $I-K$. These red
colours can be due either to an evolved, passive stellar population at
$z\ga1$, or the presence of large quantities of dust associated with
starburst activity (although edge-on discs may also contribute
significantly; Yan \& Thompson 2003). While the former population
reveals the stellar mass already assembled in galaxies as a result of
star formation at early cosmic times, the latter population reveals
the star formation at $z\sim1$ which, due to its dust-enshrouded
nature, may be missed by other surveys. Since these two populations
represent very different routes to the creation of present-day
galaxies, there is a clear desire to reliably separate them and use
them as tests of galaxy formation models.

Existing samples of EROs (for which we adopt the most common
definition of $R-K>5.3$) have been relatively small (typically no more
than about 100 objects), due to the difficulty in obtaining
near-infrared imaging data with the necessary depth and areal coverage
(e.g., Daddi et al.\ 2000; Smail et al.\ 2002; Moustakas et al.\ 2004;
Georgakakis et al.\ 2006). The arrival of infrared cameras with large
fields of view, such as WFCAM (Casali et al., in preparation), has
largely eliminated this problem, allowing samples of EROs to be
constructed which are large enough to permit further separation by
magnitude, colour, and other properties. Furthermore, by locating
infrared surveys in well-observed patches of sky, it becomes possible
to study the multi-wavelength properties of EROs and learn more about
their nature.

In this Letter, we combine near-infrared data from the UKIDSS Ultra
Deep Survey Early Data Release (Dye et al.\ 2006; Lawrence et al.\
2006) with optical data from the Subaru/\textit{XMM-Newton\/} Deep
Field (SXDF; Furusawa et al., in preparation) to construct a sample of
EROs to faint magnitudes ($K>20$) over an area larger than half a
square degree. The Ultra Deep Survey will eventually cover
0.8\,deg$^2$ to 5$\sigma$ point source limits of $J=25.0$, $H=24.0$,
and $K=23.0$ and has been sited in the SXDF because of the excellent
deep multi-wavelength data in this region. We determine photometric
redshifts for all objects and separate them into dusty star-forming
and passively-evolving galaxies on the basis of their \textit{RJK\/}
colours. We use deep radio data to determine the typical radio flux
densities and star formation rates as a function of colour and
magnitude. A more detailed analysis, including clustering and X-ray
properties, will be undertaken after the UKIDSS DR1 release, which is
expected to be a magnitude deeper than the present data, and will
provide more reliable photometry. All magnitudes in this paper are on
the Vega system. The SXDF catalogue is calibrated on the AB system,
and we use the conversion $R_{\rm Vega} = R_{\rm AB} - 0.219$ (H.\
Furusawa, private communication). We adopt a cosmology with $H_0 =
70\rm\,km\,s^{-1}\,Mpc^{-1}$ and $\Omega_{\rm m} \equiv
1-\Omega_\Lambda = 0.3$.

\section{The ERO sample}

We start with the improved version of the UDS EDR $K$-band catalogue
produced by Foucaud et al.\ (2006), and determine the completeness for
galaxies as a function of $K$-band magnitude. Galaxies with
$K\approx20$ have a median ${\rm FWHM} = 1.2''$ and so we added
objects of this size to the $K$-band WFCAM images and measured the
fraction that were recovered by the extraction process. The 50\%
completeness limit varies slightly between detectors, but averages to
$K=20.3$. We therefore eliminated all objects from the catalogue of
Foucaud et al.\ (2006) with (total) magnitudes fainter than this limit
before proceeding further. Our magnitude limit is brighter than that
of Foucaud et al.\ since they calculate the completeness for
unresolved sources. The optical data reach $R\approx<26.7$ ($5\sigma$,
$\phi2''$).

\subsection{Optical--infrared matching}

In order to facilitate matching of the near-infrared catalogue with
objects in the optical imaging, objects in various regions of sky were
next removed from the $K\leq20.3$ catalogue. Only the central region
of the UDS image, with uniform noise, was included in the matching
process. Objects were eliminated whose UDS coordinates placed them
outside the optical imaging region, or within 2\,arcsec of the edge,
to ensure that the photometric aperture of any optical match was
entirely within the image. Finally, objects which lay within the halo
or CCD bleed of a bright star were removed from the catalogue. The
remaining effective area is 2013\,arcmin$^2$.

The UDS catalogue was first matched to the catalogues of Furusawa et
al.\ (in preparation), with the closest match within 1\,arcsec being
used.  However, visual inspection of the unmatched objects revealed
that many pairs of objects remain blended in the optical catalogues,
so the UDS catalogue was instead matched to the catalogues produced by
Simpson et al.\ (2006) with a more severe deblending threshold. The
$i'$-band catalogue was used first (as the $i$-band data are the
deepest), followed by the $z'$, $R$, and $B$-band catalogues. This
reduced the number of unmatched objects by about half, to 426. These
were all inspected by eye, and the vast majority (301) were discovered
to be artifacts such as cross-talk, satellite trails, or array defects
not removed in the reduction process (see Dye et al.\ 2006). These
were deleted from the catalogue and not considered further. Most of
the remaining 125 sources are still blended in the optical catalogue,
although some lack any optical counterpart. Optical photometry was
determined for these objects directly from the Suprime-Cam images
using the UDS catalogue coordinates. A total of 443 objects have
measured $R$-band aperture magnitudes fainter than the $5\sigma$ limit
of the optical data.

In what follows, colours are measured in a 2-arcsecond aperture, while
magnitudes are total magnitudes from SExtractor's MAG\_AUTO parameter
(Bertin \& Arnouts 1996). The $K$-band image has a narrower psf than
the other images (FWHM=0.69\,arcsec compared to 0.81\,arcsec), and we
address this point below.

\subsection{Star--galaxy separation}

\begin{figure}
\begin{center}
\resizebox{0.95\hsize}{!}{\includegraphics[angle=-90]{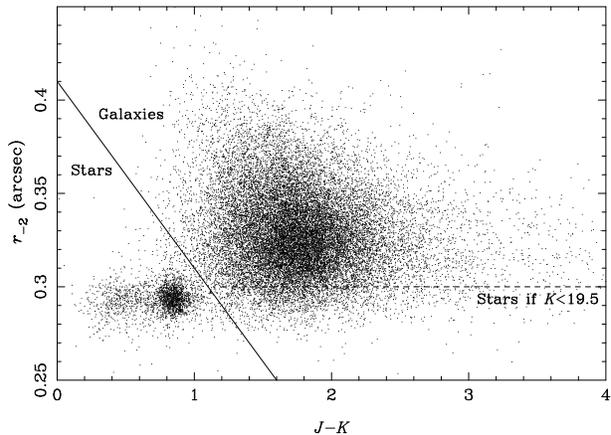}}
\end{center}
\caption[]{Plot of $r_{-2}$ (Kron 1980) against $J-K$ colour for all
objects with $K<20.3$. The solid line shows the criterion used to
separate stars from galaxies.\label{fig:kron}}
\end{figure}

We classify objects as stars or galaxies by the same method as Simpson
\& Rawlings (2002), using a combination of $J-K$ colour and Kron
(1980) radius, $r_{-2}$. There is a clear bimodality in this plot
(Fig.~\ref{fig:kron}) because galaxies rapidly become redder than
stars with increasing redshift, while those with similar colours to
stars are nearby and hence clearly extended. While there was a clear
separation between the two groups in the small sample of Simpson \&
Rawlings (2002), the much larger number of objects in the UDS EDR
means that many are scattered into `no man's land', and we determine
the locus of the minimum surface density of objects in
Fig.~\ref{fig:kron} as our separation criterion. All objects with
$r_{-2}<0.41-0.1(J-K)$ are classified as stars. However, the
combination of depth and area in the UDS will result in the detection
of a number of very late-type stars with extremely red $J-K$ colours,
and these will not be excluded by this single criterion. We therefore
also classify as stars any objects with $r_{-2}<0.3$ and $K<19.5$
(morphological classification is not possible fainter than this).

Our galaxy sample will suffer contamination from L dwarf stars, which
have $J-K>1$ (e.g., Hewett et al.\ 2006). Objects with $J-K>1.3$ and
$K>19.5$ will be classified as galaxies, and these red colours are
possessed by stars of spectral types L1--L8 (Table~9 of Knapp et al.\
2004), whose surface density to $K<14.5$ is 0.038\,deg$^{-2}$ (Reid et
al.\ 1999). If these are isotropically distributed on the sky (so the
number expected is proportional to the Euclidian volume sampled), we
expect $\sim60$ objects with $K\leq20.3$ in our survey area. These are
therefore a small contaminant.

\subsection{Number counts}

To construct our ERO sample, we first investigate the effect of the
superior $K$-band image quality by Gaussian-smoothing objects with
$R-K>5.3$ to the same FWHM as the other images and measuring the flux
loss in a 2-arcsec aperture. We find a median reduction in brightness
of 0.04\,mag, and apply this reduction to all $K$-band magnitudes in
the UDS EDR catalogue before constructing our ERO sample. In total we
find 3715 (5128; 1464) objects with $R-K>5.3$ ($R-K>5$; $R-K>6$).

\begin{figure}
\begin{center}
\resizebox{0.95\hsize}{!}{\includegraphics[angle=-90]{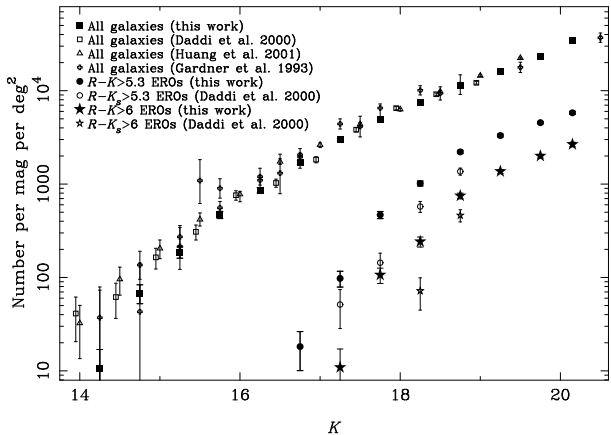}}
\end{center}
\caption[]{Differential source counts for objects in the SXDF/UDS
compared to a selection of results from the
literature.\label{fig:counts}}
\end{figure}

In Fig.~\ref{fig:counts} we compare our galaxy and ERO number counts
with determinations from the literature. Our total galaxy number
counts agree very well, except at the brightest magnitudes where there
is a slight deficit due to the fact that we have excluded regions with
very extended galaxies from our analysis. Our ERO counts are, however,
$\sim60$\,per cent higher than those of Daddi et al.\ (2000). This
could be due in part to the different $R$-band filters used in these
studies, since Daddi et al.'s filter has a long red tail with its
10\,per cent point at 8200\,\AA, while the Suprime-Cam filter
has a sharper long-wavelength cutoff with a 10\,per cent transmission
point at 7200\,\AA. Consequently, objects with a pronounced (4000-\AA)
break between these two wavelengths will appear fainter in the
Suprime-Cam images, and hence redder in $R-K$. However, we note that
we also find a much larger surface density of $R-K>5.3$ objects than
did Smail et al.\ (2002), who used the same instrument/filter
combination for their optical imaging. This may be partly due to the
small aperture we use to determine colours, since contamination from
nearby objects (which is obviously more likely for larger apertures)
will almost certainly make the aperture colour bluer. We find that the
number of EROs is decreased by 15--20\,per cent (with a larger
reduction for redder objects) if we use the colours measured in a
4-arcsecond aperture.  Finally, we note that our surface density is
similar to that of Miyazaki et al.\ (2003) who studied a small region
within the SXDF, suggesting that this region is overdense at
$z\sim1$. For objects as clustered as EROs ($r_0 \approx 12$\,Mpc;
Daddi et al.\ 2001), the effect of cosmic variance can be large, and
we estimate $\sigma_v \sim 0.2$ using the prescription of Somerville
et al.\ (2004).  We stress that the depth of the $R$-band data (the
$3\sigma$ limit in a 2-arcsecond aperture is $R\approx27.4$; Furusawa
et al., in preparation) means that photometric errors will not scatter
a significant number of the more numerous bluer objects into the ERO
class.

\section{Nature of the ERO population}

\begin{figure}
\begin{center}
\resizebox{0.95\hsize}{!}{\includegraphics{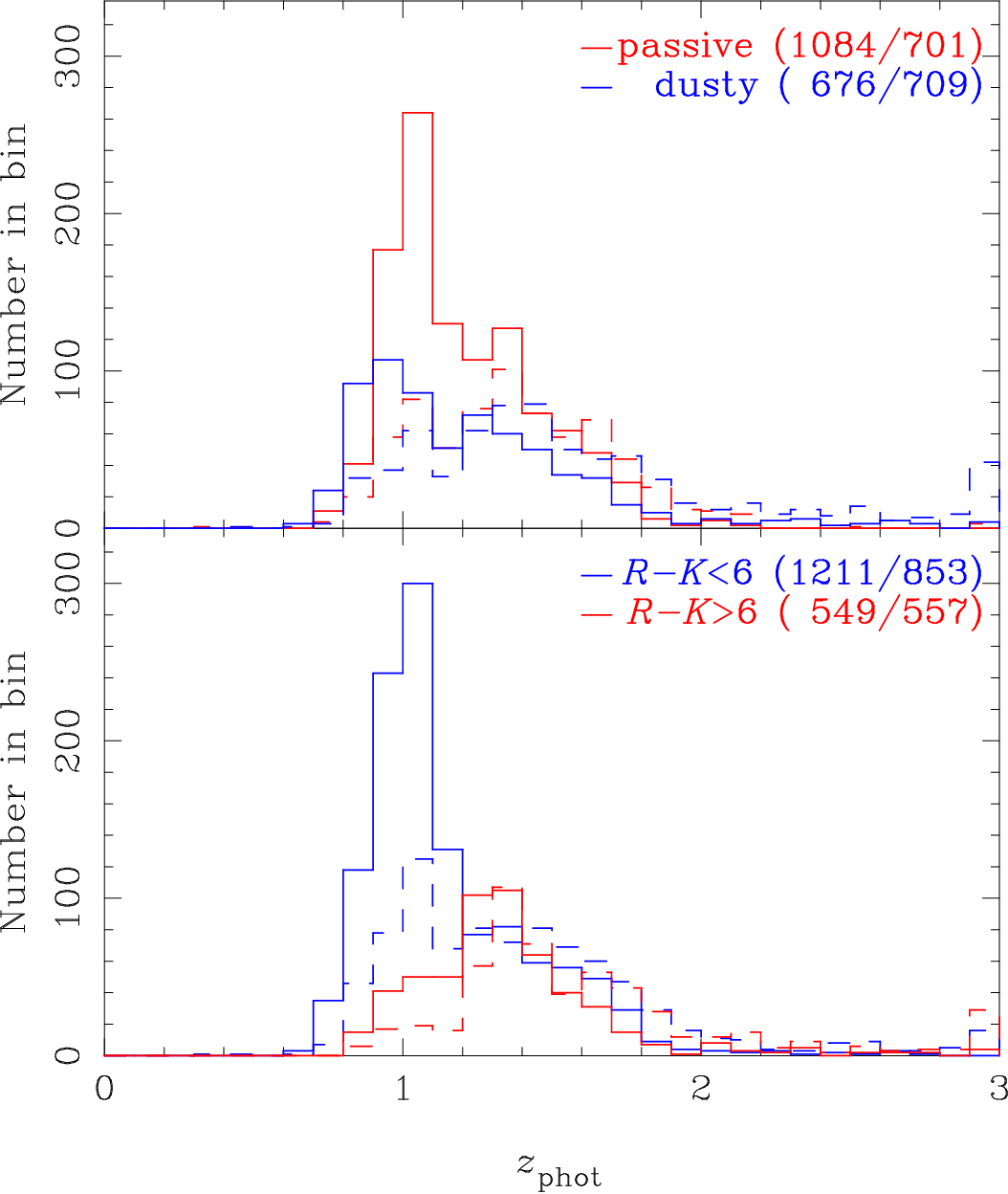}}
\end{center}
\caption[]{Histograms of the distribution of photometric redshifts,
  $z_{\rm phot}$, for subsamples of EROs. In each plot, the solid
  lines show the distribution of bright ($K\leq19.5$) EROs, while the
  dashed line show the distribution of faint ($K>19.5$) objects. The
  numbers by the line-colour key indicate the total number of bright
  and faint objects in each class, respectively.\label{fig:zhist}}
\end{figure}

Although the ERO population is composed of both passively-evolving red
galaxies and dusty starburst galaxies, there does not appear to be a
bimodality in any photometric property which allows these two classes
to be readily separated. Pozzetti \& Mannucci (2000) used $J$-band
photometry to attempt to separate the classes as heavily-reddened
objects will not show the pronounced 4000-\AA\ break apparent in the
spectra of old galaxies, and hence will have bluer $R-J$ and redder
$J-K$ colours. Smail et al.\ (2002) used a more sophisticated method
of spectral template fitting with the code \textit{HyperZ\/}
(Bolzonella, Miralles, \& Pell\'{o} 2000) and classified objects
according to whether the best-fitting model had a low or high
extinction, finding that this produced broadly the same classification
as the simpler method. Both methods rely on the accuracy of the
photometry around $\lambda_{\rm rest} \sim 4000$\,\AA, and our ability
to classify EROs is therefore limited by the quality of the $J$-band
photometry. Since the dividing line between the two classes of object
is at $J-K\approx2$ and the $J$-band photometry limit is $J=21.6$
($5\sigma$, $\phi2''$), a classification is likely to be reliable only
for the 1922 EROs with $K\leq19.5$ and we do not believe a
sophisticated analysis is warranted by the present near-infrared data
quality. We therefore opt to use Pozzetti \& Mannucci's (2000) simple
\textit{RJK\/} colour criterion to separate our sample into passive
and dusty galaxies. We find that 62\% of $K\leq19.5$ EROs are
classified as passively-evolving, although this fraction increases
when a redder colour cut is used, rising to 73\% (77\%) of objects
with $R-K>6$ ($R-K>6.5$). These fractions are broadly consistent with,
although perhaps slightly larger than, those determined from previous
studies.

We have computed photometric redshifts for all members of our ERO
sample, in order that we can derive physical quantities from observed
fluxes, by fitting synthetic galaxy templates to the $BVRi'z'JK$
photometry, plus the two short-wavelength IRAC bands (from the SWIRE
survey of Lonsdale et al.\ 2003) where available, using a modified
version of the \textit{HyperZ\/} code (see Cirasuolo
et al., in preparation, for a detailed description). We show the
results of this analysis by plotting the distribution for various
subsamples in Fig.~\ref{fig:zhist}. Objects whose best template fits
have $\chi^2>20.09$ (corresponding to an unacceptable fit at 99\,per
cent confidence) are not plotted.

Fig.~\ref{fig:zhist} also reveals the predictable results that the
reddest objects are typically at higher redshift, as are the faintest
objects. The redshift distribution for passive galaxies also appears
to be slightly narrower and more strongly peaked than that for dusty
ones. This is due to the passage of the 4000-\AA\ break through the
filters, as galaxies will only be red in $R-K$ if the break is beyond
the blue end of the $R$ filter, while the Pozzetti \& Mannucci (2000)
criterion breaks down at $z>2$ and classifies all EROs as dusty. We
caution against a more detailed analysis of the different $N(z)$
distributions given the significant difference between the photometric
accuracy of the optical and near-infrared data, even for EROs. The
deeper \textit{JK\/} data from UKIDSS DR1 will provide better
photometry, and hence a more reliable photometric analysis, for the
present sample, as well as allowing a basic study of the fainter
($K\sim21$) ERO population.

\section{Radio properties}

\begin{table*}
\caption[]{Average radio flux densities of ERO subsamples, determined
from Gaussian fitting to the variance-weighted images (the `stacking'
method; see text). For each magnitude- and colour-selected subsample,
the median photometric redshift is shown (only considering objects
with $\chi^2<20.09$), together with the number of objects stacked and
the number rejected from the stacking analysis due to being
individually detected. The mean radio flux density and star formation
rate (for stars with $M>5\rm\,M_\odot$),  adopting the median
photometric redshift for the subsample, are also
shown.\label{tab:radiofluxes}}
\begin{center}
\begin{tabular}{crc@{~~}r@{~}rr@{~~}rc@{~~}r@{~}rr@{~~}r}
\hline
Colour & \multicolumn{1}{c}{Magnitude} &
\multicolumn{5}{c}{``Passive'' EROs} & \multicolumn{5}{c}{``Dusty'' EROs} \\
threshold & \multicolumn{1}{c}{range} &
$\overline{z}_{\rm phot}$ & $N_{\rm stack}$ & $N_{\rm rej}$ &
$\overline{S}_{1.4}$ ($\umu$Jy) & SFR (M$_\odot \rm\,yr^{-1}$) &
$\overline{z}_{\rm phot}$ & $N_{\rm stack}$ & $N_{\rm rej}$ &
$\overline{S}_{1.4}$ ($\umu$Jy) & SFR (M$_\odot \rm\,yr^{-1}$) \\
\hline
& $K\leq18.5$ &
1.05 & 220 & 54 & $6.7\pm1.3$ & $8.5\pm1.7$ &
0.96 & 109 & 55 & $15.0\pm1.0$ & $15.3\pm1.0$ \\
$R-K>5.3$ & $18.5<K\leq19.5$ &
1.20 & 819 & 104 & $4.2\pm0.6$ & $7.2\pm1.1$ &
1.25 & 463 & 98 & $10.3\pm0.5$ & $19.7\pm0.9$ \\
& $19.5<K\leq20.3$ &
1.37 & 832 & 78 & $3.4\pm0.4$ & $8.1\pm0.9$ &
1.46 & 800 & 83 & $5.3\pm0.7$ & $14.5\pm2.0$ \\
\hline
& $K\leq18.5$ &
1.24 & 64 & 12 & $9.4\pm1.7$ & $17.6\pm3.1$ &
1.18 & 14 & 8 & $11.7\pm1.9$ & $19.5\pm3.2$ \\
$R-K>6.0$ & $18.5<K\leq19.5$ &
1.36 & 351 & 55 & $5.3\pm0.4$ & $12.3\pm0.9$ &
1.32 & 131 & 29 & $9.4\pm0.7$ & $19.9\pm1.5$ \\
& $19.5<K\leq20.3$ &
1.44 & 414 & 34 & $3.1\pm0.6$ & $8.2\pm1.5$ &
1.60 & 321 & 31 & $4.4\pm1.2$ & $15.0\pm4.0$ \\
\hline
& $K\leq18.5$ &
1.35 & 19 & 2 & $12.1\pm2.0$ & $27.7\pm4.6$ &
1.53 & 6 & 1 & $10.5\pm4.8^\dagger$ & $32.3\pm14.7$ \\
$R-K>6.5$ & $18.5<K\leq19.5$ &
1.47 & 131 & 25 & $5.8\pm0.7$ & $16.3\pm1.9$ &
1.68 & 40 & 7 & $10.2\pm2.0$ & $39.2\pm7.6$ \\
& $19.5<K\leq20.3$ &
1.64 & 177 & 16 & $3.4\pm0.8$ & $12.2\pm2.7$ &
1.81 & 124 & 10 & $4.8\pm1.2$ & $21.6\pm5.5$ \\
\hline
\multicolumn{10}{l}{$^\dagger$Ellipse parameters fixed during fit due
to high noise.}
\end{tabular}
\end{center}
\end{table*}

Simpson et al.\ (2006) have presented a 1.4-GHz radio image of the
SXDF taken with the NRAO Very Large Array which reaches an rms noise
level of 12\,$\umu$Jy\,beam$^{-1}$ in the central regions. At the
median redshift of our sample, $z\approx1.3$, a star formation rate
(SFR) for $M>5\rm\,M_\odot$ stars of $10\rm\,M_\odot\,yr^{-1}$
produces a flux density of 5\,$\umu$Jy (Condon 1992). Although this is
obviously below the detection threshold for any individual source, we
can make statistical detections of relatively modest SFRs in samples
of $\ga100$ objects since the $5''\times4''$ beam will encompass all
radio emission from the host galaxy.  We note that starburst galaxies
will be unresolved at the resolution of this radio image.

We study the mean radio properties of our sample in two ways. First,
we produce small postage stamps from the radio map of Simpson et al.\
(2006) at our objects' positions and sum these after weighting each by
the inverse of the pixel noise variance at its position. We exclude
all objects where any of the nine pixels (the scale is
1.25\,arcsec\,pixel$^{-1}$) in the radio image closest to the source
position has a signal-to-noise ratio in excess of 3, so that the mean
is not biased by a few objects associated with bright radio
sources. We then fit an elliptical Gaussian profile to this stacked
image using Ivan Busko's \texttt{n2gaussfit} task in \textsc{iraf},
and calculating the flux density from the fit using the prescription
of Condon (1997). For the second method, we construct a histogram of
pixel values from the radio map at the positions of objects in our
sample, and fit a Gaussian to the pixels after they have been
iteratively sigma-clipped to exclude values more than two standard
deviations from the mean, thus automatically removing bright
outliers. We refer to these two methods as the stacking and histogram
methods, respectively. Since the histogram method fits a Gaussian to
the convolution of true source fluxes with the near-Gaussian pixel
noise distribution, it is most reliable if the source flux density
distribution is either itself Gaussian, or has a dispersion much less
than that of the overall pixel noise. We prefer to use the image
stacking method to derive mean radio flux densities, but confirm that
the results from the histogram method are consistent in all cases.

\begin{figure}
\begin{center}
\resizebox{0.95\hsize}{!}{\includegraphics[angle=-90]{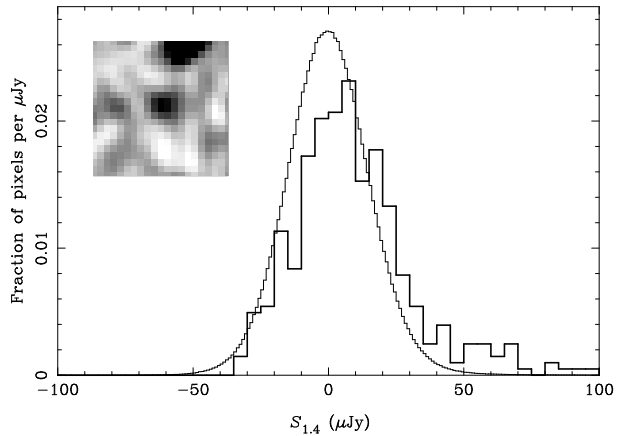}}
\end{center}
\caption[]{Radio properties of EROs with $18.5<K\leq19.5$, $R-K>6$,
and classified as `passive'. The histogram shows the radio image pixel
values at the locations of the sample objects (thick line), compared
to the pixel values of the 2013\,arcmin$^2$ over which EROs were
selected (thin line). The inset shows the results of the image
stacking, after rejecting individually detected sources. A clear
positive signal can be seen in both
analyses.\label{fig:stack_example}}
\end{figure}

To determine the likelihood of spurious detections in our image
stacks, we ran Monte-Carlo tests by producing 1000 lists of between
100 and 1000 sources each located at random positions within the
unmasked region of the SXDF/UDS images. The mean flux density at the
positions in each list was then calculated by both the stacking and
histogram methods. Since the stacked radio images are essentially
signal-free, the Gaussian fitting technique was highly unstable and
instead we measured the signal in a 5$\times5$ pixel box in the centre
of the image and corrected this for the 20\,per cent flux loss
expected for the beam size. Both methods produced a mean flux density
of $-0.2\,\umu$Jy, and we have therefore corrected our measured flux
densities for this offset. The standard deviations on the mean for a
sample of 100 sources were 0.9 and 2.1\,$\umu$Jy for the stacking and
histogram methods, respectively, and reduced with sample size $N$
approximately as $N^{-1/2}$.

Table~\ref{tab:radiofluxes} shows the results of the stacking analysis
for 18 ERO subsamples (separated by magnitude, colour, and
classification), while Fig.~\ref{fig:stack_example} shows the stacking
and histogram analyses for one subsample, indicating the reality of
the statistical detection. From our analysis, galaxies classified as
`dusty' typically have twice the mean radio flux density of `passive'
galaxies, and the fraction of individually-detected objects is about
twice as high (except in the faintest magnitude bin, where we have
argued that the classification is unreliable). Smail et al.\ (2002)
also find that most EROs with flux densities of a few tens of $\umu$Jy
are dusty. However, while a smaller fraction of passive galaxies are
detected individually, those that are tend to be brighter: in the
$18.5<K\leq19.5$ bin, 28 (13) of 104 individually-detected passive
EROs have $S_{1.4}>100\,\umu$Jy ($S_{1.4}>300\,\umu$Jy), while only 10
(4) of 98 dusty EROs are as bright (for reference, the Fanaroff \&
Riley (1974) break is at $S_{1.4}\approx6$\,mJy at $z=1.3$). This
suggests that a few per cent of the passive ERO population may be
active galaxies (cf.\ Alexander et al.\ 2002), although such objects
can also be found in the dusty population (e.g., EDXS N2\_21; Willott
et al.\ 2003). Among the dusty population, we find that the SFR is
uncorrelated with magnitude, but may be correlated with colour.

Our mean radio flux densities are consistent with those determined by
Georgakakis et al.\ (2006), although our larger sample size produces
much smaller uncertainties.  We find that dusty galaxies typically
have about twice the 1.4-GHz flux density of passive galaxies with the
same colours and magnitudes, and hence have about twice the star
formation rate since the median redshifts of identically-selected
passive and dusty EROs are similar (in computing the median redshifts
for our subsamples, we ignore cases where the best-fitting photometric
redshift has $\chi^2>20.09$).  However, the true difference in star
formation rates is likely to be larger than this since the
individually detected dusty objects are likely to be the high end of a
broad distribution of star formation rates whereas the individually
detected passive objects are likely to be AGNs. As a final point, we
note that the sample of 5834 non-ERO galaxies with $18.5<K\leq19.5$
and $R-K\leq5.3$ has 440 objects detected at $>3\sigma$ significance,
while the remaining sources have a mean flux density of $S_{1.4} =
5.3\pm0.2\,\umu$Jy, consistent with the value derived for the passive
ERO population in this magnitude range.

\section{Summary}

We have described the construction of a sample of extremely red
objects (EROs) within the Subaru/\textit{XMM-Newton\/} Deep Field
(SXDF) and UKIDSS Ultra Deep Survey (UDS) which is much larger than
previous samples. We have measured a higher surface density than found
in other fields, but comparable to that measured in a small sub-region
of our field, suggesting the SXDF samples a cosmic overdensity at
$z\sim1$. After using \textit{RJK\/} colours to separate the sample
into passively-evolving (62\% of EROs with $K\leq19.5$) and dusty
star-forming galaxies, we have used a deep radio map to study the
radio properties of these two populations. Both populations are
detected at high significance in a stacked radio image, with the dusty
sample having a mean flux density (and derived mean star formation
rate) equal to approximately twice that of the passive sample. The
mean radio flux density of the passive sample is consistent with that
of non-ERO galaxies with the same magnitudes.

\section*{Acknowledgments}

This paper is partially based on data collected at Subaru Telescope,
which is operated by the National Astronomical Observatory of
Japan. The United Kingdom Infrared Telescope is operated by the Joint
Astronomy Centre on behalf of the U.K. Particle Physics and Astronomy
Research Council (PPARC). The National Radio Astronomy Observatory is
a facility of the National Science Foundation operated under
cooperative agreement by Associated Universities, Inc. We thank the
staff of these facilities for making these observations possible, and
the anonymous referee for helpful suggestions. We gratefully
acknowledge financial support from the PPARC (CS, MC, SF, SR) and the
Royal Society (OA, IS).

%\bsp

\label{lastpage}

\end{document}